\documentstyle[12pt,epsf]{article}
\author{A.HASEGAWA, Y.IWASAKI, K.SAKAMOTO, N.NODA,\\H.KOUNO and M.NAKANO$^*$\\
Department of Physics, Saga University
\\
$^*$University of Occupational and Environmental Health}
\title{Antiproton Production in p+d Reaction \\
at Subthreshold Energies}
\date{}

\begin{document}
\maketitle
\begin{abstract}
 An enhancement of antiprotons produced in $p+d$
reaction in comparison
 with ones in $p+p$ elementary reaction is investigated.
 In the neighborhood of
 subthreshold energy the enhancement is caused by the difference of
 available energies for antiproton production. The cross section in $p+d$
 reaction, on the other hand, becomes just twice of the one in elementary
 $p+p$ reaction at the incident energy far from
 the threshold energy when non-nucleonic components in deuteron target
 are not considered.
\end{abstract}

Antiproton productions have been one of the most interesting subthreshold
production process since they have been observed in proton-nucleus
reactions more than thirty years ago [1-3]. 

About a decade ago, the production of antiprotons at subthreshold in
nucleus-nucleus was observed [4-6] and various models have been proposed
to explain experimental data, models based on the assumption of kinetic
and chemical equilibrium [7-9], and models in terms of multiple
interactions [10,11].
First fully relativistic transport calculations for antiproton
production including antiproton annihilation as well as the change
of the quasi-particle properties in the medium have shown that the
antiproton yields for $p+A$ and $A+A$ can be well reproduced simultaneously
when employing proper self energies for the baryons in the dense
medium (the relativistic BUU approach)[12-16].

Five years ago the problem was taken up at KEK[17] aiming to carry out
measurements of subthreshold antiproton productions with light
ions such as $d$ and $\alpha$ . In particular, the cross section ratio of
antiproton productions between proton induced and deuteron induced
reactions is insensitive to uncertainties of antiproton reabsorption
in a target nucleus and of the antiproton production cross
section in elementary NN interaction near the subthrshold,
which give large ambiguity in any theoretical model calculation.
The most surprising finding was an enormous enhancement of
antiproton productions in $d+A$ reaction at lower incident
energies per nucleon and the existence of non-nucleon components
in the deuteron wave function was suggested to cure the situation.

To explore this suggestion the BUU approach group performed calculations
employing the deuteron wave function from which had been fitted to
deuteron fragmentation data at high energy and noted that the
present data do not yet provide clear evidence for non-nucleon
components in the deuteron wave function[18].

%****************************************************
% figure 1
%****************************************************

\begin{figure}[htbp]
  \epsfxsize=5cm
  \centerline{\epsfbox{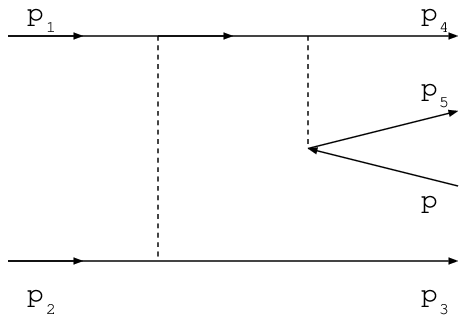}}  
\label{fig1}
\caption{The diagram for antiproton producing mechanism in proton-proton
collision.}
\end{figure}

%***************************************************

In this note we take up the most simple reaction $p+d \rightarrow \bar{p}$ and
clarify the
enhancement of antiproton productions in comparison with the elementary
process when we do not consider non-nucleonic components in the deuteron
wave function.  First, we treat the elementary $\bar{p}$ production case, $
p+p \rightarrow \bar{p}ppp$.
The Feynman diagram is shown in Fig.1 where the four-dimensional momenta
of particles are also indicated. The total cross section is described
as follows,

%******************************************************
% equation (1)
%******************************************************

\begin{eqnarray}
\sigma \; = \; \frac{1}{4 \sqrt{{(p_1 \cdot p_2)}^2 - {m_1}^2 {m_2}^2}}
\frac{1}{{(2 \pi) }^8} \int \frac{d \mbox{\boldmath $p$}_3}{2E_3}
\frac{d \mbox{\boldmath $p$}_4}{2E_4}
\frac{d \mbox{\boldmath $p$}_5}{2E_5}
\frac{d \mbox{\boldmath $p$}}{2E}
{\mid T_{pp \rightarrow \bar{p}ppp} \mid}^2 \nonumber \\
\times
\delta (p_1+p_2-p_3-p_4-p_5-p) 
\end{eqnarray}

%******************************************************

where $m_i$ $(i=1 \sim 5)$ and mass of antiproton are proton mass $M$, and
$E_i \;and \;E$ are 
on shell proton and antiproton energies respectively.
The absolute square of invariant amplitude \\
${\mid T_{pp \rightarrow \bar{p}ppp} \mid}^2$ is given

%****************************************************
% equation (2)
%*****************************************************

\begin{equation}
{\mid T_{pp \rightarrow \bar{p}ppp} \mid}^2 \; \cong \;
T_{pp}^2 \cdot (2M) \cdot (2M)
\end{equation}

%***************************************************

We assume that the cross section of antiproton production in low incident
energies near the threshold energy is approximately proportion to the
four body phase space. Then, we take a constant value for the absolute
square of invariant amplitude which is fitted to data of antiproton
production in the neighborhood of the threshold energy as shown in Fig.2.
The obtained constant value of
${\mid T_{pp \rightarrow \bar{p}ppp} \mid}^2$ can be
separated
into two
factors , the incident proton part emitting antiprotons $T_{pp}^2$ and the
target proton part $(2M) \cdot (2M)$ which approximates the square of
normalization
factor, ($2E_2$ ) $\cdot$ ($2E_3$ ),
of the wave functions before and after the reaction.
This $T_{pp}^2$ is inserted into the incident proton
part in an absolute square of invariant amplitude
$T_{pd \rightarrow \bar{p}}$ which
describes $\bar{p}$
 productions in pd interaction.

%****************************************************
% figure 2
%****************************************************

\begin{figure}[htbp]
  \epsfxsize=8cm
  \centerline{\epsfbox{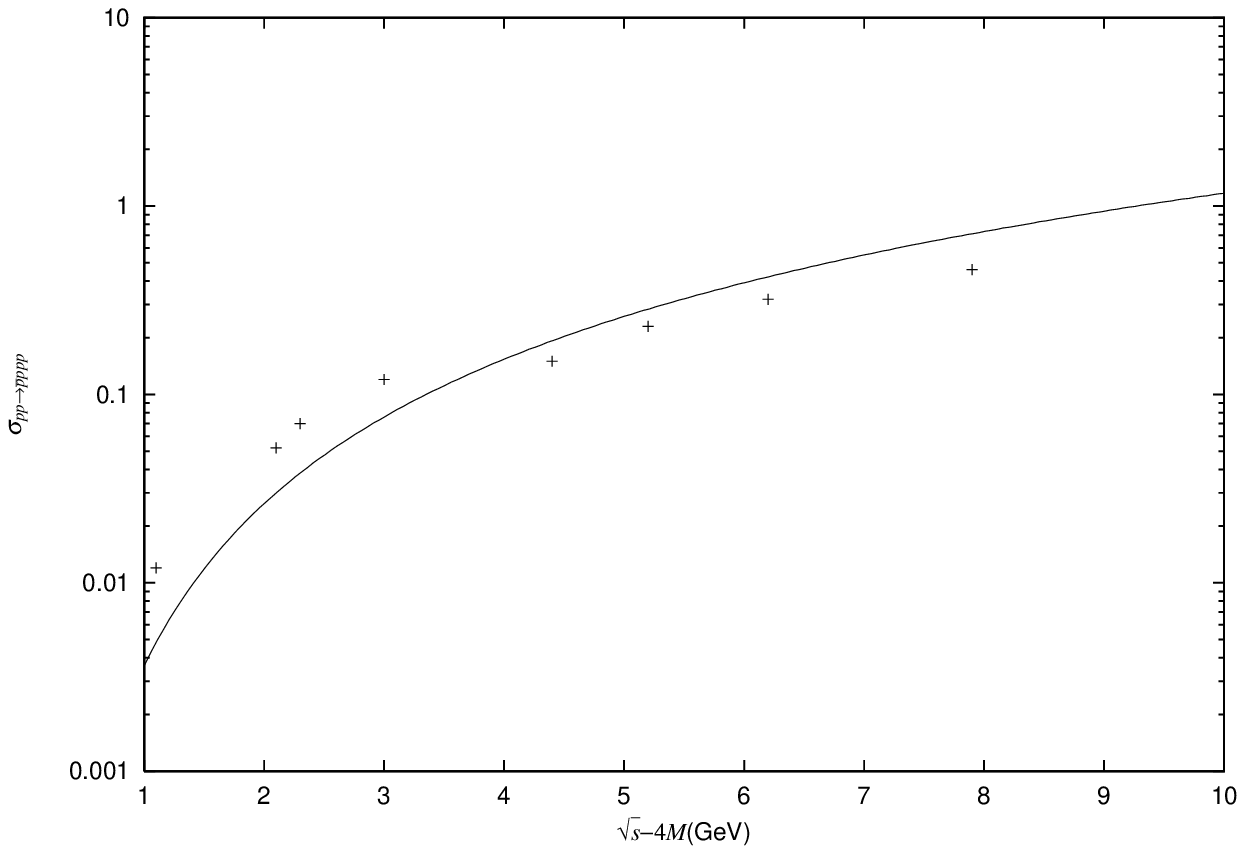}}  
\label{fig2}
\caption{The cross section for antiproton production in proton-proton
collisions,circles and squares refered from ref.19 and ref.20,respectively.
The line indicates our parameterization of the cross section as described
in the text.}
\end{figure}

%***************************************************

Next, we proceed to reactions $p+d \rightarrow \bar{p}ppd$
and $p+d \rightarrow
\bar{p}pppn$ of which Feynman
diagrams are shown in Fig.3. Both of diagrams show that a fast proton
collides with a deuteron at rest and products an antiproton while the
target deuteron keeps a bound state or breaks up in two nucleons.
In this note we treat the internal motion of deuteron non-relativisticaly
and use the deuteron wave function which is solved with the Reid soft core
potential by the superposition of a few Gaussians [19].

%*************************************************
% equation (3)
%************************************************

\begin{equation}
\tilde{\phi}_d( \mbox{\boldmath $p$}) \;=\;
\sum_{i=1}^3 a_i e^{- \mbox{\boldmath {\footnotesize$p$}}^2/d_i^2}
\end{equation}

where the parameters $a_i$ and $d_i$ are given as

\begin{eqnarray*}
&a_1  =  540(GeV)^{-3/2}, \qquad   &d_1 = 0.058(GeV), \\
&a_2  = 72 , \qquad &d_2 = 0.158, \\
&a_3  = 4.4. \qquad &d_3 = 0.58.
\end{eqnarray*} 

%***********************************************

Applying this wave function to the deuteron target in pd collision
the observed energy spectra of backward proton are well reproduced [20].
The absolute square of amplitude of ${\mid T_{pd \rightarrow \bar{p}ppd}
\mid}^2$
is given

%************************************************
% equation (4)
%************************************************

\begin{equation}
{\mid T_{pd \rightarrow \bar{p}ppd} \mid}^2 \; \cong \; T_{pp}^2 \cdot
(2m_d) \cdot (2m_d) \cdot {(f_a({(\mbox{\boldmath $p$}_3-
\mbox{\boldmath $p$}_2)}^2))}^2
\end{equation}

%************************************************
\noindent
where $m_d$  denotes the mass of deuteron and
${(f_a({(\mbox{\boldmath $p$}_3-\mbox{\boldmath $p$}_2)}^2))}^2$
is a probability
that the internal motion in the target deuteron absorbs a momentum
$\mbox{\boldmath $p$}_3-\mbox{\boldmath $p$}_2$
transferrd from the incident proton,

\begin{equation}
f_a(\mbox{\boldmath $q$}^2)\;=\;\frac{1}{{(2\pi)}^2}
\int d \mbox{\boldmath$p$} \prime
\phi(\mbox{\boldmath $p$} \prime)
\phi(\mbox{\boldmath $p$} \prime - \frac{1}{2} \mbox{\boldmath $q$})
+
\phi(\mbox{\boldmath $p$} \prime + \frac{1}{2} \mbox{\boldmath $q$})]
\end{equation}
\[
\mbox{\boldmath $q$}=\mbox{\boldmath $p$}_3
-
\mbox{\boldmath $p$}_2
\]

Also, we obtain ${\mid T_{pd \rightarrow \bar{p}pppn} \mid}^2$ of the deuteron
breakup process as follows

\begin{equation}
{\mid T_{pd \rightarrow \bar{p}pppn} \mid}^2
\; \cong \; T_{pp}^2 \cdot (2m_d) \cdot {(2M)}^2
(f_b{(\mbox{\boldmath $p$}_2; \mbox{\boldmath $p$}_3,
\mbox{\boldmath $p$}_6))}^2
\end{equation}

\begin{equation}
f_b(\mbox{\boldmath $p$}_2;\mbox{\boldmath $p$}_3,\mbox{\boldmath $p$}_6)
\;=\;
\phi(\mbox{\boldmath $p$}_3-\frac{1}{2}\mbox{\boldmath $p$}_2)+
\phi(\mbox{\boldmath $p$}_6-\frac{1}{2}\mbox{\boldmath $p$}_2)
\end{equation}

where
${(f_b(\mbox{\boldmath $p$}_2;\mbox{\boldmath $p$}_3,
\mbox{\boldmath $p$}_6))}^2$ 
is proportion to a probability that either nucleon
in target deuteron accepts a momentum transferrd from the incident proton.
Thus, we obtain the invariant cross section of $\bar{p}$-productions
in $p+d$ reaction,

\begin{eqnarray}
\sigma \; = \; \frac{1}{4 \sqrt{{(p_1 \cdot p_2)}^2 - {m_1}^2 {m_2}^2}}
\frac{1}{{(2 \pi) }^8}
\mbox{\Large [ }
T^2_{pp} \cdot (2m_d) \cdot (2m_d) \nonumber \\
\times
\int \frac{d \mbox{\boldmath $p$}_3}{2E_3}
\frac{d \mbox{\boldmath $p$}_4}{2E_4}
\frac{d \mbox{\boldmath $p$}_5}{2E_5}
\frac{d \mbox{\boldmath $p$}}{2E}
{(f_a({(\mbox{\boldmath $p$}_3-\mbox{\boldmath $p$}_2)}^2))}^2 
\delta (p_1+p_2-p_3-p_4-p_5-p) \nonumber \\
+T_{pp}^2 \cdot 2m_d \cdot {(2M)}^2
\int
\frac{1}{{(2\pi)}^3}
\frac{d \mbox{\boldmath $p$}_3}{2E_3}
\frac{d \mbox{\boldmath $p$}_4}{2E_4}
\frac{d \mbox{\boldmath $p$}_5}{2E_5}
\frac{d \mbox{\boldmath $p$}_6}{2E_6}
\frac{d \mbox{\boldmath $p$}}{2E}
{(f_b({(\mbox{\boldmath $p$}_2;\mbox{\boldmath $p$}_3,
\mbox{\boldmath $p$}_6)}^2))}^2 \nonumber \\
\times
\delta (p_1+p_2-p_3-p_4-p_5-p_6-p)
\mbox{\LARGE  ]} \nonumber \\
\; 
\end{eqnarray}

The four-momentum of deuteron is defined as
$p_2 = (E_2 - \epsilon, \mbox{\boldmath $p$}_2 )$ where $E_2$ 
and $\epsilon$
are the center of mass energy and the relative binding energy respectively.

\vspace{5mm}
%****************************************************
% figure 3
%****************************************************

\begin{figure}[htbp]
  \epsfxsize=12cm
  \centerline{\epsfbox{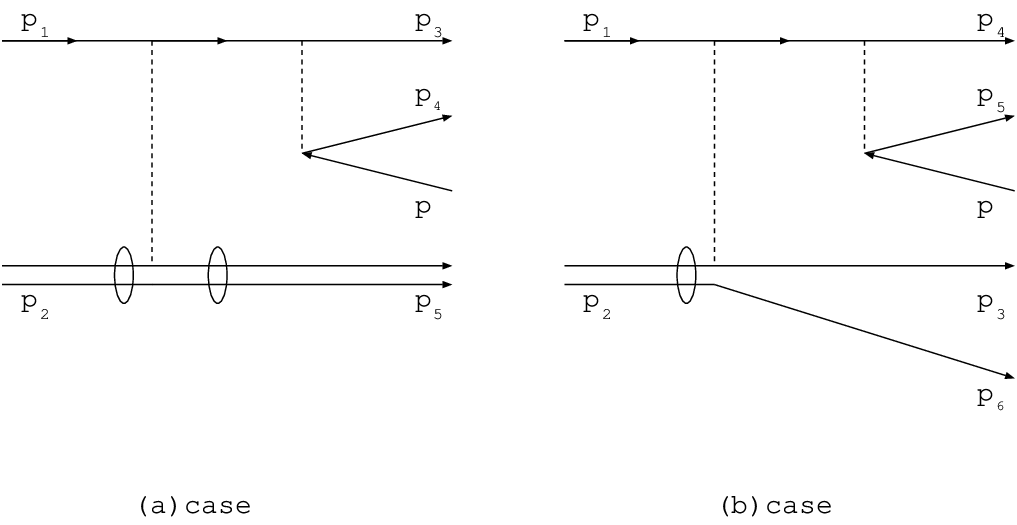}}  
\label{fig3}
\caption{The diagram for antiproton producing mechanism in proton-deuteron
collision. The projectile proton collides with one of the two nucleons
in deuteron and produces antiprotons in two processes (a)
$P+d \rightarrow \bar{p}ppd$ (deuteron bound) and (b)
$p+d \rightarrow \bar{p}pppn$(deuteron breakup).}
\end{figure}

%***************************************************

The figure 4 shows the ratio of total cross sections of $p+p$ and $p+d$
reactions as a function of the incident energy. The available energies
for incident energy are $7M$ and $5M$ + $\epsilon$ for $p+p$ and $p+d$ reactions
respectively.The latter is the subthreshold energy for p production
in pd interaction.
We note here that the available energy of incident deuteron energy is also
$5M$ + $\epsilon$ per nucleon when the deuteron collides with the
proton target at rest.

%****************************************************
% figure 4
%****************************************************

\begin{figure}[htbp]
  \epsfxsize=8cm
  \centerline{\epsfbox{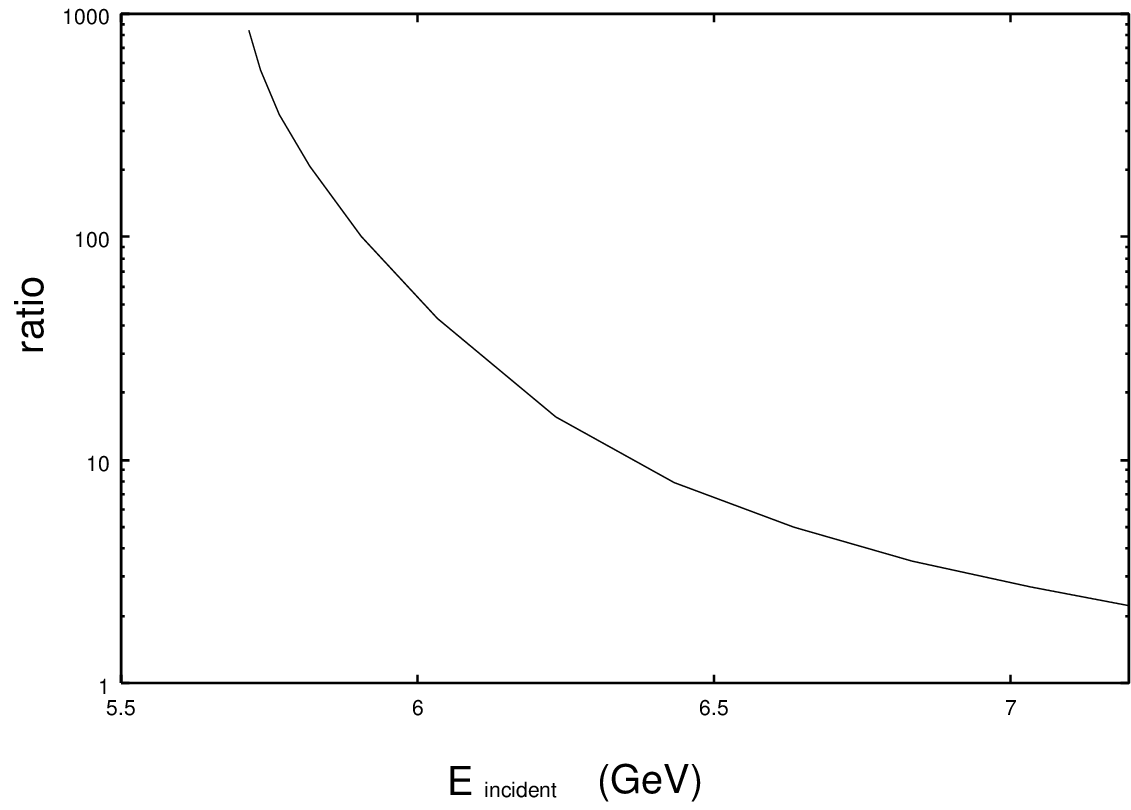}}  
\label{fig4}
\caption{The ratio of antiproton productions in pp and pd collisions as
a function of the incident energy.}
\end{figure}

%***************************************************

At low incident energies in the neighborhood of threshold of
$p+p \rightarrow \bar{p}$ reaction, the enormous enhancement of
the ratio of
antiprotons produced in $p+d$ reaction and ones in $p+p$ reaction 
is reasonably caused due to the
difference of the available energies for $\bar{p}$ productions.
The figure 5 shows two processes in $p+d$ reaction, the deuteron bound process
and the deuteron breakup process. The deuteron breakup process produces
more antiprotons than the bound process since the former has a five-body
phase space larger than a four-body phase space in the latter process and
since the transition matrix of the former process includes an internal
wave function of deuteron while the one of the latter process has an
absolute square of wave function of deuteron.

%****************************************************
% figure 5
%****************************************************

\begin{figure}[htbp]
  \epsfxsize=8cm
  \centerline{\epsfbox{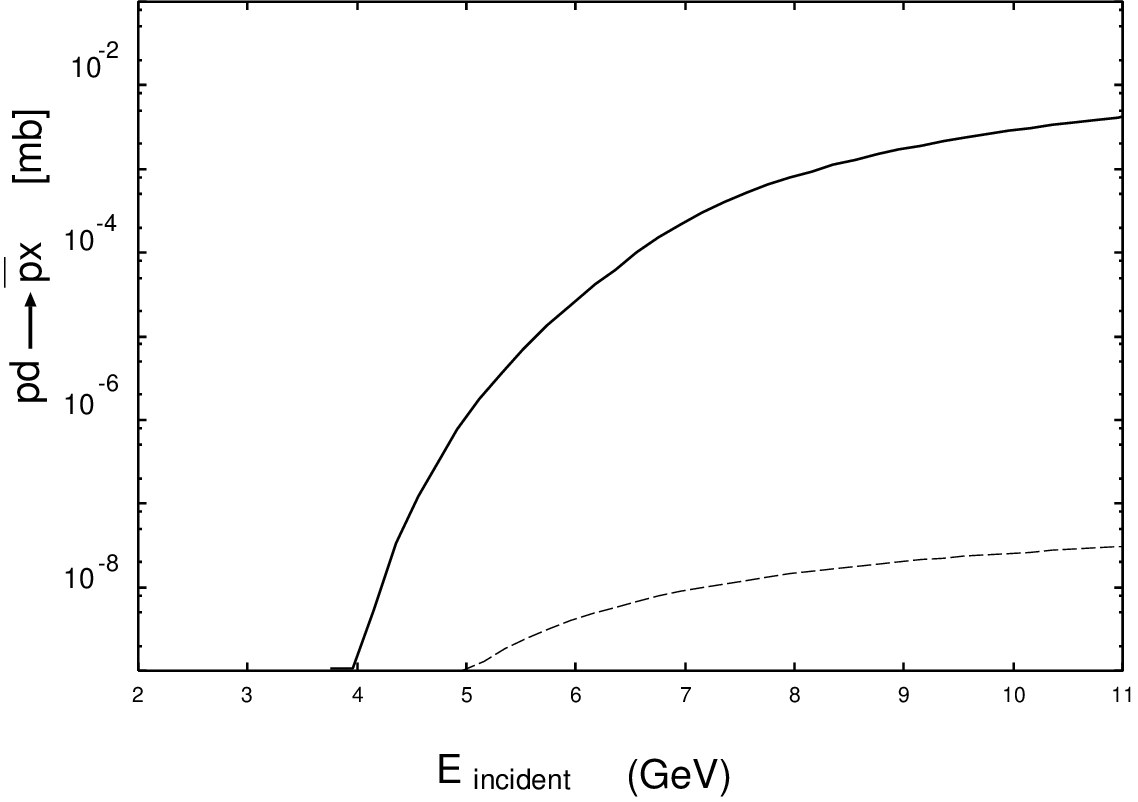 }}  
\label{fig5}
\caption{The cross sections as a function of the incident energy.
The solid line and the dashed line indicate the deuteron breakup process
and the deuteron bound process, respectively.}
\end{figure}
%***************************************************

We take notice of the cross section of the deuteron breakup process, the
second term in equation (8). The cross section takes the maximum value
when two broken-up nucleons have respectively a half of momentum of
deuteron before the collision with proton. In the laboratory
system $(\mbox{\boldmath $p$}_2=0,E=m_d)$,
if we replace the two internal wave functions
in Eq.(8) with $\delta$ -functions,

\begin{equation}
\phi( \mbox{ \boldmath $p$}_3) \rightarrow {(2\pi)}^ \frac{3}{2}
\delta ( \mbox{\boldmath $p$}_3) \quad and \quad 
\phi( \mbox{ \boldmath $p$}_6) \rightarrow {(2\pi)}^ \frac{3}{2}
\delta ( \mbox{ \boldmath $p$}_6)
\end{equation}

the cross section is rewritten as follows,

%***********************************************************
% equation (10)
%***********************************************************
\begin{eqnarray}
&&\sigma \; =  \; \frac{1}{4} \frac{1}{v_{rel}}
\frac{1}{E_1} \frac{1}{{(2 \pi)}^8}
\cdot T_{pp}^2 \cdot 4M \nonumber \\
&& \mbox{\Large [ } 
\int \frac{d \mbox{\boldmath $p$}_3}{2E_3}
\frac{d \mbox{\boldmath $p$}_4}{2E_4}
\frac{d \mbox{\boldmath $p$}_5}{2E_5}
\frac{d \mbox{\boldmath $p$}}{2E} 
\delta (E_1+m_d-M-E_3-E_4-E_5-E) \nonumber \\
&& \times \delta (\mbox{\boldmath $p$}_1-\mbox{\boldmath $p$}_3
-\mbox{\boldmath $p$}_4
-\mbox{\boldmath $p$}_5-\mbox{\boldmath $p$}) \nonumber \\
&& + \int 
\frac{d \mbox{\boldmath $p$}_6}{2E_6}
\frac{d \mbox{\boldmath $p$}_4}{2E_4}
\frac{d \mbox{\boldmath $p$}_5}{2E_5}
\frac{d \mbox{\boldmath $p$}}{2E}
\delta (E_1+m_d-M-E_6-E_4-E_5-E) \nonumber \\
&& \times \delta (\mbox{\boldmath $p$}_1-\mbox{\boldmath $p$}_6
-\mbox{\boldmath $p$}_4
-\mbox{\boldmath $p$}_5-\mbox{\boldmath $p$})
\mbox{\LARGE  ]}  
\end{eqnarray}

%**************************************************************

where $v_{rel}$ denotes the relative velocity between incident
proton and target
deuteron. This cross section is just twice of the one in the elementary
process if the small binding energy of deuteron is not taken into account.
The interference term disappears, since this term corresponds to a deuteron
spectator process and the energy-momentum conservation for residual four
nucleons can not be satisfied because four nucleons are on shell. Then,
the ratio of cross sections approaches to a factor two as the incident
energy goes up far away from the threshold energy. We are going to
calculate the differential cross section including non-nucleonic
components in deuteron internal state.

The authors would like to thank T. Kohmura, T. Maruyama and S. Nakamura
for useful discussions and making many valuable comments.
\\
\noindent
References:\\
\noindent
1) O. Chamberlain et al., Nuovo Ciment{\bf 3}(1956),447.\\
2) T. Elioff et al., Phys. Rev.
{\bf 128}(1962),869.\\
3) D. Dorfan et al., Phys. Rev. Lett. {\bf 14}(1965),995.\\
4) A. A. Baldin et al., JETP Lett. {\bf 47}(1988),137.\\
5) J. B. Carrol et al., Phys. Rev. Lett. {\bf 62}(1989)1829.\\
6) A. Shor et al., Phys. Rev. Lett. {\bf 63}(1989),2192.\\
7) P. Koch and C. B. Dover, Phys. Rev. {\bf C40}(1989),145.\\
8) C. M. Ko and X. Ge, Phys. Lett. {\bf B205}(1988),195.\\
9) C. M. Ko and L. H. Xia, Phys. Rev. {\bf C40}(1989),R1118.\\
10) A. Shor, V. Perez-mendez and K. Ganezer, Nucl. Phys. {\bf A514}
(1991),717.\\
11) P. Danielewicz, Phys. Rev. {\bf C42}(1990),1564.\\
12) K. Weber, B. Blattel, W. Cassing,  H. C. Donges, V. Koch,
A. Lang and U.       Mosel, Nucl. Phys. {\bf A539}(1992),713.\\
13) T. Maruyama, B. Blattel, A. Lang, W. Cassing, U. Mosel
and K. Weber, Phys.     Rev. Lett. {\bf B297}(1992),228.\\
14) G. Batko, W. Cassing, U. Mosel, K. Niita and
Gy. Wolf, Phys. Lett. {\bf B256}        (1991),331.\\
15) S. W. Huang, Guo-qiang Li, T. Maruyama and A.
Faessler, Nucl. Phys. {\bf A547}      (1992),653.\\
16) S. Teis, W. Cassing, T. Maruyama and U. Mosel, Phys. Lett. {\bf B319}
(1993),47.\\
17) J. Chiba et al., Nucl. Phys. {\bf A553}(1993),771C.\\
18) W. Cassing, G. Lykasov and S. Teis, Z. Phys. {\bf A348}(1994),247.\\
19) M. Antinucci et al., Lett. Nuovo Cimento 6(1973),121.\\
20) C. S. Shen and G. B. Berkey, Phys. Rev. 171(1968),1344.\\
21) R. V. Reid, Ann. of Phys.(n.Y.){\bf 50}(1968),411.\\
22) Y. Haneisi and T. Fujita, Phys. Rev. {\bf C33}(1986),260.\\

\end{document}